\documentstyle[emulateapj,psfig]{article}

\lefthead{Brotherton et al.}
\righthead{Binary Quasar}

\begin{document}

\title{Discovery of a Radio-loud/Radio-quiet Binary Quasar\altaffilmark{1}}
\author{M. S. Brotherton\altaffilmark{2}, Michael D. Gregg\altaffilmark{2,3}, R. H. Becker\altaffilmark{2,3}, S. A. Laurent-Muehleisen\altaffilmark{2,3}, \\
R. L. White\altaffilmark{4}, S.~A. Stanford\altaffilmark{2,3,5}}

\altaffiltext{1}{Based on observations at the W. M. Keck Observatory.}
\altaffiltext{2}{Institute of Geophysics and Planetary Physics, Lawrence Livermore National Laboratory, 7000 East Avenue, P.O. Box 808, L413, Livermore, CA 94550}
\altaffiltext{3}{Physics Dept., University of California--Davis}
\altaffiltext{4}{Space Telescope Science Institute, 3700 San Martin Dr., 
Baltimore, MD 21218}
\altaffiltext{5}{Visiting Astronomer at the Infrared Telescope Facility
(IRTF), which is operated by the University of Hawaii under contract with
the National Aeronautics and Space Administration.}

\begin{abstract}

We report the discovery of a small separation quasar pair 
($z$ = 0.586, $O$ = 18.4, 19.2, sep. = 2$\farcs$3) associated with the 
radio source FIRST J164311.3+315618 ($S_{1400} = 120$ mJy).  
The spectrum of the brighter quasar (A) has a much stronger narrow 
emission-line spectrum than the other (B), and also stronger
Balmer lines relative to the continuum.
The continuum ratio of the spectra is flat in the blue 
($\lambda_{obs} < 6000$ \AA) at about 2.1, 
but falls to 1.5 at longer wavelengths.  A $K^{\prime}$ image
shows two unresolved sources with a flux ratio of 1.3.
The different colors appear to result from the contribution of 
the host galaxy of B, which is evident from Ca II and high-order Balmer 
absorption lines indicative of a substantial young stellar population.
New 3.6 cm VLA observations show that the compact radio source 
is coincident with quasar A (B is only marginally detected).  
We rule out the lensing hypothesis because the optical flux ratio is A/B 
$\approx$ 1.5--2 while the radio flux ratio is A/B $\gtrsim40$, 
and conclude that this system is a binary.
Moreover, the radio-loud quasar is a compact steep spectrum source.
FIRST J164311.3+315618A, B is the lowest redshift and smallest separation 
binary quasar yet identified.

\end{abstract}
\section{Introduction}

Small separation quasar pairs at similar redshifts are closely scrutinized
as gravitational lens candidates.  While dozens of lensed quasars have
been confirmed, wide separation (3\arcsec-10\arcsec) high-redshift
$O^2$ pairs (both quasars are radio faint) are more problematic.
Djorgovski et al. (1987) discovered PKS 1145-071, the first $O^2R$ pair
(one quasar is radio bright and the other is radio faint):
PKS 1145-071 must be a binary because lensing cannot produce two images with
extremely different flux ratios in the optical and the radio.
The existence of $O^2R$ quasar binaries, plus the fact that only some
10\% of optically selected quasars are radio-loud, lead Kochanek, Falco, \&
Mu\~noz (1997) to make a statistical argument that most of the 
wide-separation $O^2$ lens candidates are actually binary quasars.

Understanding the frequency of binary quasars is important for interpreting 
lensing statistics.  Such systems also provide clues to the 
origins or fueling of quasar activity, and why strong radio jets are found in
some objects but not others.  If galaxy interactions and mergers
activate quasars (e.g., Sanders et al. 1988; Hernquist 1989; Mihos \& Hernquist
1996), then the rapid evolution of the space density of 
quasars, and its peak during the epoch $z = 2-3$, may reflect the 
merger history of galaxies.  If galaxies are assembled hierarchically, 
then quasars may trace galaxy evolution and star formation (e.g., Barnes 1998). 

In a spectroscopic search for quasars detected by the FIRST survey 
(Faint Images of the Radio Sky at Twenty centimeters; Becker et al. 1995),
we discovered a radio-loud, radio-quiet binary quasar system associated 
with the radio source FIRST J164311.3+315618 (hereafter FIRST J1643+3156).
In this $Letter$, we describe our observations of this system and compare it
to other confirmed binary quasars.  
We adopt $H_0 = 50$ km s$^{-1}$ Mpc$^{-1}$, $q_0$ = 0.
With $z=0.586$ and a separation of $2\farcs3$ (20 kpc), 
FIRST J1643+3156A, B is the nearest and smallest separation binary quasar known.

\section{Observational Results}

We identified FIRST J1643+3156 as a quasar candidate based on its
radio emission ($S_{1400} = 120$ mJy, integrated flux density) 
in the FIRST survey, its stellar classification by the APM 
(e.g., McMahon \& Irwin 1992) on both the blue
($O$ = 18.0 mag.) and red ($E$ = 18.0 mag.) POSS plates, 
and its blue color ($O-E = 0$).  
While the current limit of the FIRST Bright Quasar Survey (Gregg et al. 1996; 
White et al. 1999) is $E=17.8$, we are observing fainter targets as part of
the FIRST Faint Quasar Survey (Becker et al. 1998).

We observed FIRST J1643+3156 on 15 July 1998 (UT) with the Keck II telescope
using the Low Resolution Imaging Spectrometer (LRIS; Oke et al. 1995).  On the
acquisition camera we noticed that there were two stellar objects separated
by 2$\arcsec$\ near the target coordinates.
We obtained exposures of three minutes on the brighter
southern component (A) and ten minutes on the fainter northern component (B).
The 300 line mm$^{-1}$ grating with a 1$\arcsec$\ slit
gave a resolution $\leq$ 10 \AA. 
We employed standard data reduction techniques within the NOAO IRAF
package, and used a standard star to divide out atmospheric features.
The redshift of each is 0.586 based on [O II] $\lambda$3727 and
[O III] $\lambda$5007, with a difference at the $\sim$100 km s$^{-1}$
level.  The broad Mg II $\lambda$2800 profiles are similar
with FWHM = 3400$\pm$300 km s$^{-1}$, but the broad Balmer lines and 
narrow-line spectra differ.
Quasar B also shows high-order Balmer and Ca II H and K absorption lines,
and has a redder continuum (Fig. 1).

We took a $K^{\prime}$ image using NSFCAM at the IRTF on
22 August 1998, when conditions were photometric with $0\farcs8$\ seeing.
The total on-source exposure time was 960~s.  After linearization and
flatfielding, the data were sky-subtracted, registered, 
and summed using DIMSUM\altaffilmark{1}.  Standard stars from Persson et al. 
(1998) were used to yield magnitudes on the CIT system.
Component A has $K^{\prime}$=16.6, B has $K^{\prime}$=17.0,
and both are unresolved.

\altaffiltext{1}{DIMSUM is the Deep Infrared Mosaicing Software package,
developed by P.\ Eisenhardt, M.\ Dickinson, A.\ Stanford, and J.\ Ward,
which is available as a contributed package in IRAF.}

The FIRST survey (resolution 5$^{\prime\prime}$) is insufficient to resolve
the radio source or to determine definitively whether it is associated with 
component A or B. 
We obtained a $0\farcs7$ resolution map on August 31, 1998 (UT)
with the NRAO\altaffilmark{2} Very Large Array (VLA)
in the B configuration, at 3.6 cm, with 14 minutes exposure time. 
Data were calibrated and reduced in the standard way using
the AIPS analysis package.  
The resulting cleaned map resolves a radio source of 
small angular extent ($\approx1\farcs4$, PA$=$71$^{\circ}$).  The peak and total
flux densities are 16 mJy and 36 mJy, respectively.

We used the 9 nearest HST Guide Stars to obtain an astrometric solution on the
Second Generation Digital Sky Survey with an RMS of $\sim 0\farcs1$ (GSC 1.2 
has a positional accuracy of $0\farcs3 - 0\farcs4$ per star).  This solution
was transferred to the IRTF image using 3 unresolved objects which
appeared in both bands.  The final $1\sigma$ absolute positional error
is $\sim 0\farcs3$.  Figure~2 shows the centroid positions of
components A and B (crosses) on the VLA 3.6 cm contour map.
Quasar~A is coincident with the compact radio source while Quasar~B is 
associated with a marginally detected (0.4 mJy, $4\sigma$) radio source.
At 3.6 cm the peak-to-peak radio flux ratio is S$_{\rm A}$/S$_{\rm B}$$=$40.

\altaffiltext{2}{The National Radio Astronomy Observatory is a facility of
the National Science Foundation operated under cooperative agreement by
Associated Universities, Inc.}

FIRST J1643$+$3156 has been detected in other radio surveys, including:
Westerbork Northern Sky Survey 
(393$\pm$47 mJy at 325 MHz; Rengelink et al. 1997),
Texas (407$\pm$29 mJy at 365 MHz; Douglas et al. 1996), and the
NRAO VLA Sky Survey (121$\pm$3.7 mJy at 1.4 GHz; Condon et al. 1998).
The radio spectrum is steep with a spectral index (325 MHz to 8 GHz)
of $\alpha=-0.73$ (S$_{\rm \nu}\propto\nu^{\alpha}$).

A search of the ROSAT public archive revealed that FIRST J1643$+$3156 appears
serendipitously in a PSPC observation.
The source is clearly detected (5 ksec effective exposure time), 
but lies near a support rib and 
outside the inner support ring (offaxis angle$=$31.6$^{\prime}$) where
the point response function is degraded.
We extraced the source using a circular aperture of 2.5$^{\prime}$,
subtracted an appropriate background, and corrected for vignetting.
We used XSPEC 
to fit the background subtracted spectrum with an absorbed powerlaw using a 
fixed value of N$_{\sl H}$$=$2.2$\times$10$^{20}$ cm$^{-2}$ (Dickey \& Lockman 
1990).  The fit proved adequate ($\chi_{\nu}^2 = 1.05$, 13 dof) 
and yielded a photon index of 2.7 $\pm$ 0.2 and a 0.1-2.4 keV flux of
1.1$\times$10$^{-12}$ erg s$^{-1}$ cm$^{-2}$ 
(L$_x$ = 1.1$\times10^{45}$ ergs s$^{-1}$).

Table 1 summarizes the positions and separable properties of the 
FIRST J1643+3156 system.  A correction for the low Galactic reddening of 
A$_B$ = 0.07 has not been applied.
Given the semi-arbitrary absolute magnitude that separates Seyfert galaxies
from quasars ($M_B = -23$, for $H_0=50, q_0 = 0$, Veron-Cetty \& Veron 1998),
A is a quasar and B is a Seyfert 1 galaxy.  Below, we 
will continue to refer to both components as quasars.

\section{Discussion}

We have discovered a binary quasar system.
We can rule out the lensing hypothesis for FIRST J1643+3156A,B 
because while the quasars have similar optical magnitudes, one is radio-loud
while the other is radio-quiet.  Additionally,
their optical colors and spectra differ significantly, with the brighter
quasar possessing a stronger narrow-line spectrum and stronger Balmer
emission lines relative to the continuum.  Finally, we see no evidence 
for a lensing galaxy in our $K^{\prime}$\ image; our detection limit
is $K^{\prime} \approx 18$ (coincident with the quasars), 
while the expected magnitude for a lensing galaxy
causing a 2$\farcs$3 separation is $\sim$14-15 magnitude (e.g.,
Keeton et al. 1997; Jackson et al. 1998).

The color difference of the two quasars appears to be
primarily the result of a host galaxy in B. The spectral division in
the bottom panel of Fig. 1 shows an approximately constant A/B $\approx$ 2.1
in the blue, which drops to about 1.5 at longer wavelengths.
In the $K^{\prime}$\ band the ratio has dropped further to
1.3.  The colors and shape of the spectral division are consistent with 
a Balmer jump from a stellar contribution to the spectrum of B, which we 
might expect to see given its Seyfert-level luminosity.  

Figure 3 shows Ca II and high-order Balmer absorption lines
characteristic of a young ($<$0.5 Gyr) stellar population 
(at the system redshift and not intervening).
If the AGN in A and B have the same optical-infrared spectral energy
distribution, then the stellar light of B contributes $\geq 30\%$ of the flux 
at $R$ and $\geq 60\%$ of the flux at $K^{\prime}$.  
We can estimate the spectrum of the host
galaxy of quasar~B under the above assumption.   We
clipped the narrow emission lines from the quasar~A spectrum and then
scaled it so that the flux in Mg II matched that of quasar~B.  
Subtracting this spectrum from that of quasar~B results
in a recognizable galaxy spectrum (Fig. 3, bottom)
consistent with that of a starburst galaxy (dotted line).
The starburst in the quasar B host galaxy may have been triggered by
the interaction with quasar A.

Of additional interest is that the radio emission associated with quasar A
is compact ($\theta \sim 1.4\arcsec = 6.2 h^{-1}$ kpc) 
and has a steep spectrum ($\alpha$ = $-$0.73), 
indicating that A is a compact steep spectrum (CSS) source.  
Radio-loud quasars are
also X-ray loud compared to radio-quiet quasars, and the more
luminous quasar A very likely emits the majority of the observed X-rays.
Quasar A has radio, optical (especially the strong narrow-emission line
spectrum), and X-ray properties consistent with those of 
other CSS sources (as reviewed by O'Dea 1998).
One hypothesis is that CSS sources are ``frustrated'' radio jets, unable
to drill through a surrounding dense medium.
The other leading hypothesis is that CSS sources are the
progenitors of large-scale classical double sources.  
Disturbed morphologies or interactions, including tidal tails, are
ubiquitous among the host galaxies of CSS sources (Gelderman 1994, 1996),
and perhaps such an interaction (as in the present example of this 
binary quasar) is required to ignite a powerful radio-loud AGN.

There is a small difference in the quasars' redshifts.
A cross-correlation analysis indicates that A is redshifted
relative to B by 282$\pm$98 km s$^{-1}$.
The narrow-line peak wavelengths suggest
a smaller value: [O III] $\lambda$5007 and [O II] $\lambda$3727 in A are
redshifted by 80$\pm$10 km s$^{-1}$ relative to the same lines in B.
Our spectrum of A was obtained with a 1\arcsec\ east-west slit,
close to the radio position angle, and shows that
the [O III] $\lambda$5007 emission is spatially extended 
by $\approx5\arcsec$ (after subtracting in quadrature the continuum 
extent).  CSS sources as a class display this ``alignment effect''
where optical and radio emission have coincident position angles and related
features (Gelderman 1994, 1996; De Vries et al 1997). 
The aligned emission in CSS sources appears to be dominated by emission
lines (De Vries et al. 1998), which are very luminous in these objects,
and may be generated by shock-ionized gas from the interaction of the 
radio jet with the gaseous environment of the host galaxy 
(Bicknell et al. 1997).

Assuming that our binary is bound and isolated, 
we can calculate a limit to the total system mass:
\begin{equation}
M_A + M_B \geq \frac{Rv^2}{2G}
\end{equation}
\noindent
$M_A$ and $M_B$ are the masses of the quasars, $R$\ is their projected
separation (10.1 $h^{-1}$\ kiloparsecs), 
$v$\ is their radial velocity difference, and $G$\ is the
Gravitational constant.  The total mass is greater than 
7.7$\times10^{9} h^{-1} M_{\sun}$ for a velocity difference of 80 km s$^{-1}$,
which is a very modest limit.

Kochanek, Falco, \& Mu\~noz (1998) review the known quasar pairs 
with the purpose of discriminating between gravitationally lensed and binary 
systems.  They find three quasar pairs with greatly discrepant optical and 
radio flux ratios ($O^2R$ pairs):
the aforementioned PKS~1145--071 (Djorgovski et al. 1987),
MGC~2214+3550 (Mu\~noz et al. 1998), and Q~1343+2640 (Crampton et al. 1988).  
FIRST J1643+3156 brings the total to four known $O^2R$ binary quasar systems.

The properties of FIRST J1634+3156 are remarkably similar to those of 
MGC 2214+3550 ($z=0.88, I=18.8, 19.3,$ sep. = $3\farcs0$).
The brighter quasar in MGC 2214+3550 is again a CSS radio source.
Other similarities include: optical luminosities,
optical A/B flux ratios, radio luminosities, radio spectra
($\alpha = -0.86$ for MGC 2214+3550), radio angular size,
physical separation, radio position angle relative to the binary, and an
alignment effect (based on a Hubble Space Telescope 
NICMOS $H$-band image of MGC 2214+3550, available from the CASTLe Survey 
web page, Kochanek et al. 1999).  
Some of these may be selection biases;
still, given the small number of systems so far identified,
the similarities suggest that there may be a small range of parameters for
galaxy interactions that lead to radio-loud/radio-quiet binary systems.
 
\section{Conclusions}

We have identified a new binary quasar associated with the radio
source FIRST J164311.3+315618.  The system is
binary and not lensed because one quasar is radio loud 
while the other is radio quiet, and there are other inconsistencies with the
lensing hypothesis.  The brighter quasar is a compact steep spectrum source
with a strong narrow emission-line spectrum and extended [\ion{O}{3}]
$\lambda$5007 along the radio axis.  The spectrum of the fainter quasar, 
formally a Seyfert 1 galaxy, has a significant stellar contribution from a 
young population as evidenced by high-order Balmer absorption lines.
As the lowest redshift ($z=0.586$) binary quasar known, 
FIRST J164311.3+315618A,B provides the best available laboratory for 
investigations of the binary quasar phenomenon, and for investigations of what 
conditions may lead to radio-loud activity as opposed to radio-quiet activity.

\acknowledgments

We thank Dan Stern, Carlos De Breuck, and the staffs of the VLA, IRTF, and
Keck Observatory for their assistance.  
This research has made use of the NASA/IPAC Extragalactic Database (NED) which
is operated by the Jet Propulsion Laboratory, California Institute of 
Technology, under contract with the National Aeronautics and Space 
Administration.  The W. M. Keck Observatory is a scientific partnership 
between the University of California and the California Institute of Technology,
made possible by the generous gift of the W. M. Keck Foundation.
We acknowledge support from the National Science Foundation under grants
AST-94-19906 and AST-94-21178.
This work has been performed under the auspices of the U.S. Department of Energy
by Lawrence Livermore National Laboratory under Contract W-7405-ENG-48.


\vfil\eject

\def\s{$\,$}
\def\m{$-$}
\def\p{$+$}
\def\e{$\pm$}
\begin{deluxetable}{lcccccccc}
\tablewidth{0pt}
\tablenum{1}
\tablecaption{Binary Positions and Properties} 
\tablehead{Component& R.A. & Decl. & $z$\tablenotemark{a} & $E$ & $O-E$ & $K^{\prime}$ & $S_{3.6_{cm}}$ & $M_B$\tablenotemark{b} \nl
 & (J2000) & (J2000) & & & & & ($mJy$) & }
\startdata
A & 16 43 11.33 & +31 56 18.3 & 0.5867& 18.5& $-$0.1 & 16.6 & 36.4 & $-$23.3 \nl
B & 16 43 11.38 & +31 56 20.6 & 0.5862& 19.0& 0.2 & 17.0 & $0.4\tablenotemark{c}$ & $-$22.8 \nl
\enddata
\tablenotetext{a}{Calculated based solely on [O III] $\lambda$5007 wavelengths.}
\tablenotetext{b}{Assuming $H_0 = 50, q_0 = 0$\ and computed for direct
comparison with Veron-Cetty \& Veron (1998).}
\tablenotetext{c}{4 $\sigma$\ detection.}
\end{deluxetable}

\clearpage
\centerline{
\psfig{file=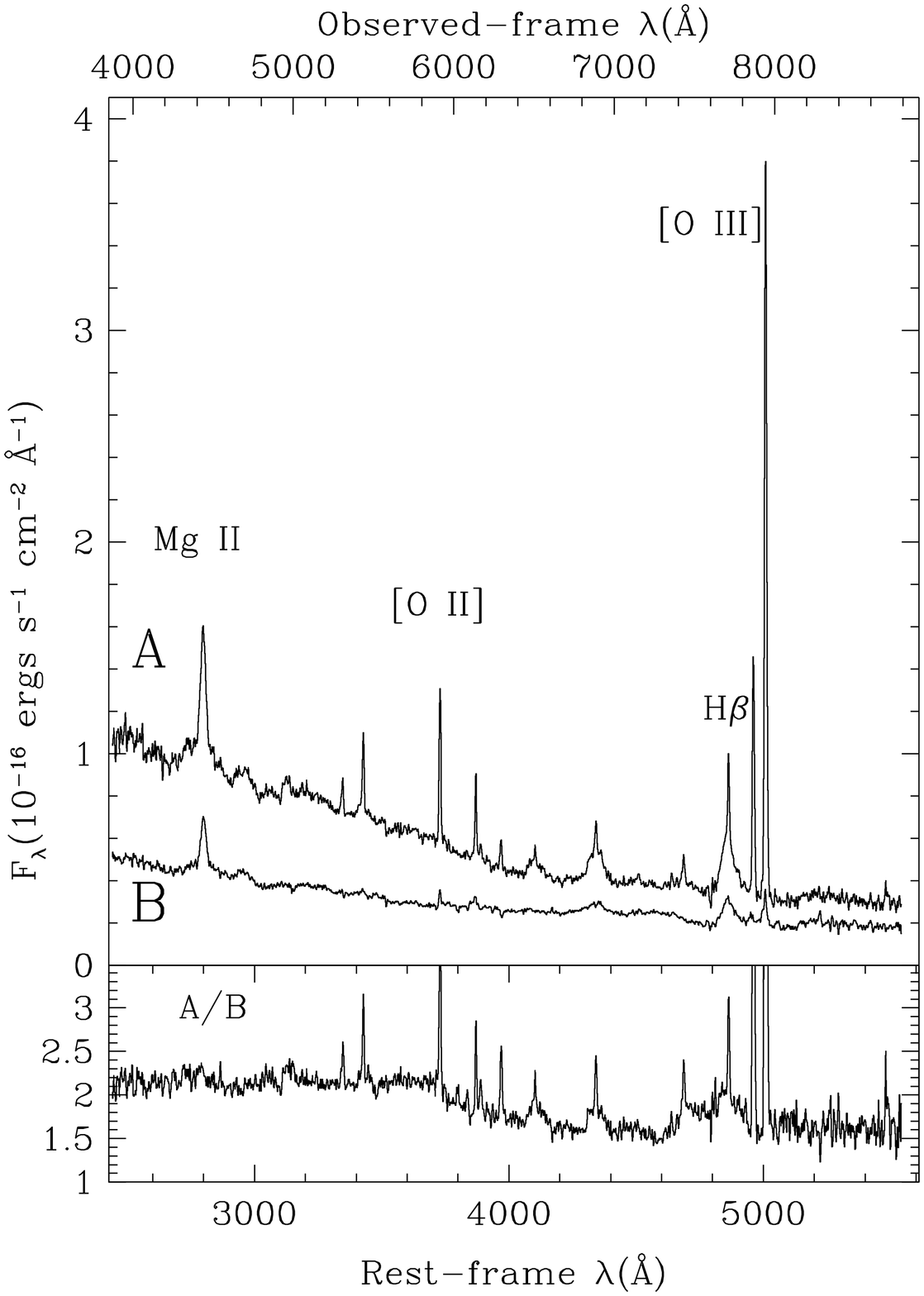,height=18cm}}
\figcaption{Upper panel: Keck II LRIS spectra of the binary component A
(3 min. exposure) and B (10 minute exposure).  The spectra have been boxcar
smoothed by three pixels. The lower panel shows the ratio of A to B.  Rest-frame
wavelengths assume $z=0.586$.}

\centerline{
\psfig{file=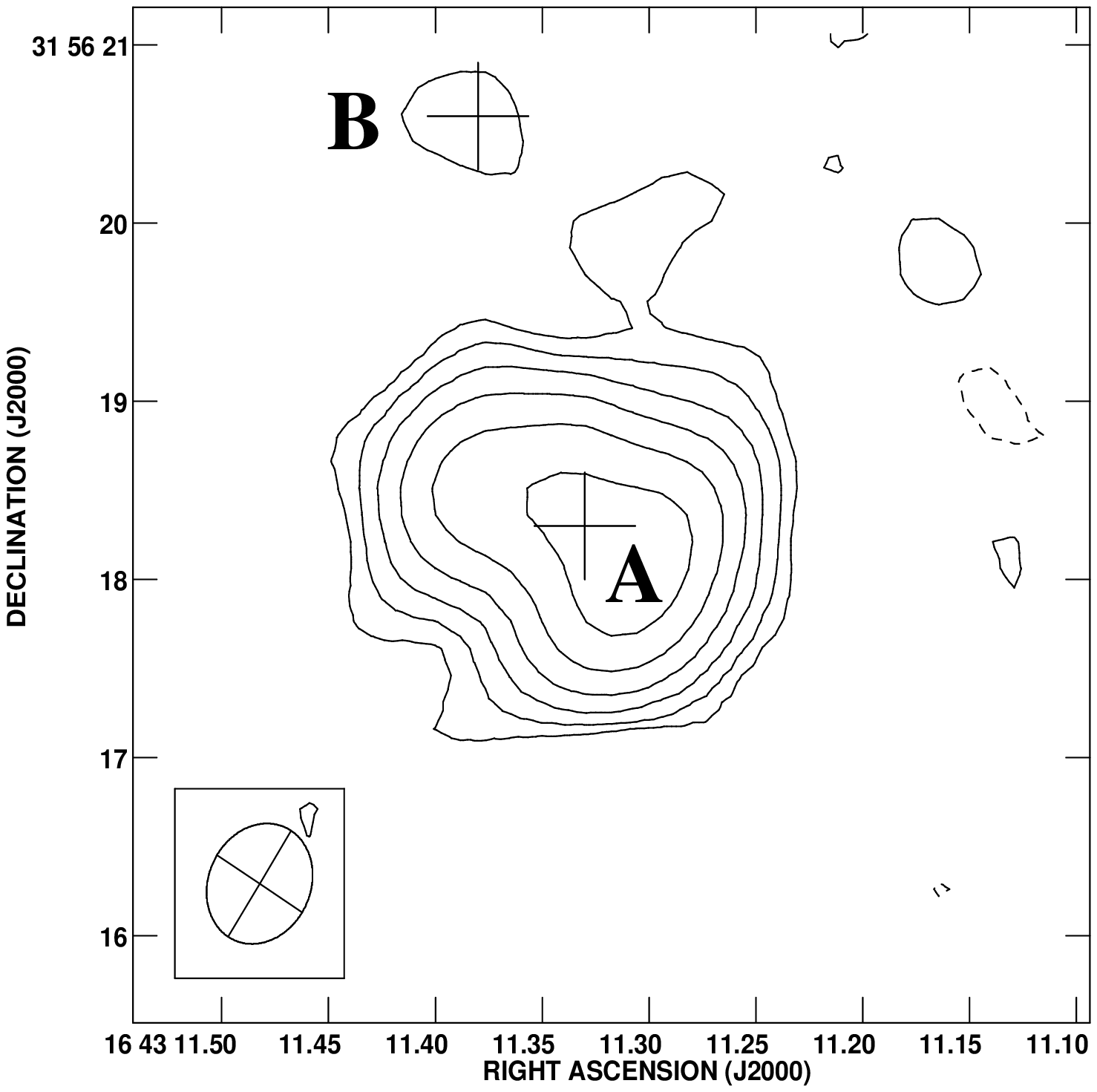,angle=-90}}
\figcaption{
Total intensity 3.6 cm radio countour map of FIRST J1643$+$3156.  
Peak flux is 16 mJy and contour levels are at $-$0.25, 0.25, 0.5, 1, 2,
4, and 8 mJy.  The integrated flux of Component A is 36 mJy and its largest
angular size is 1$\farcs$4.  The map RMS is 0.11 mJy and source B is
detected at the $\sim$4$\sigma$ level.  The optical/infrared positions are
marked and the size of the crosses indicate the formal astrometric
uncertainty.  The scale is 4.4$h^{-1}$ kpc arcsec$^{-1}$ for $q_o$$=$0.
}

\centerline{
\psfig{file=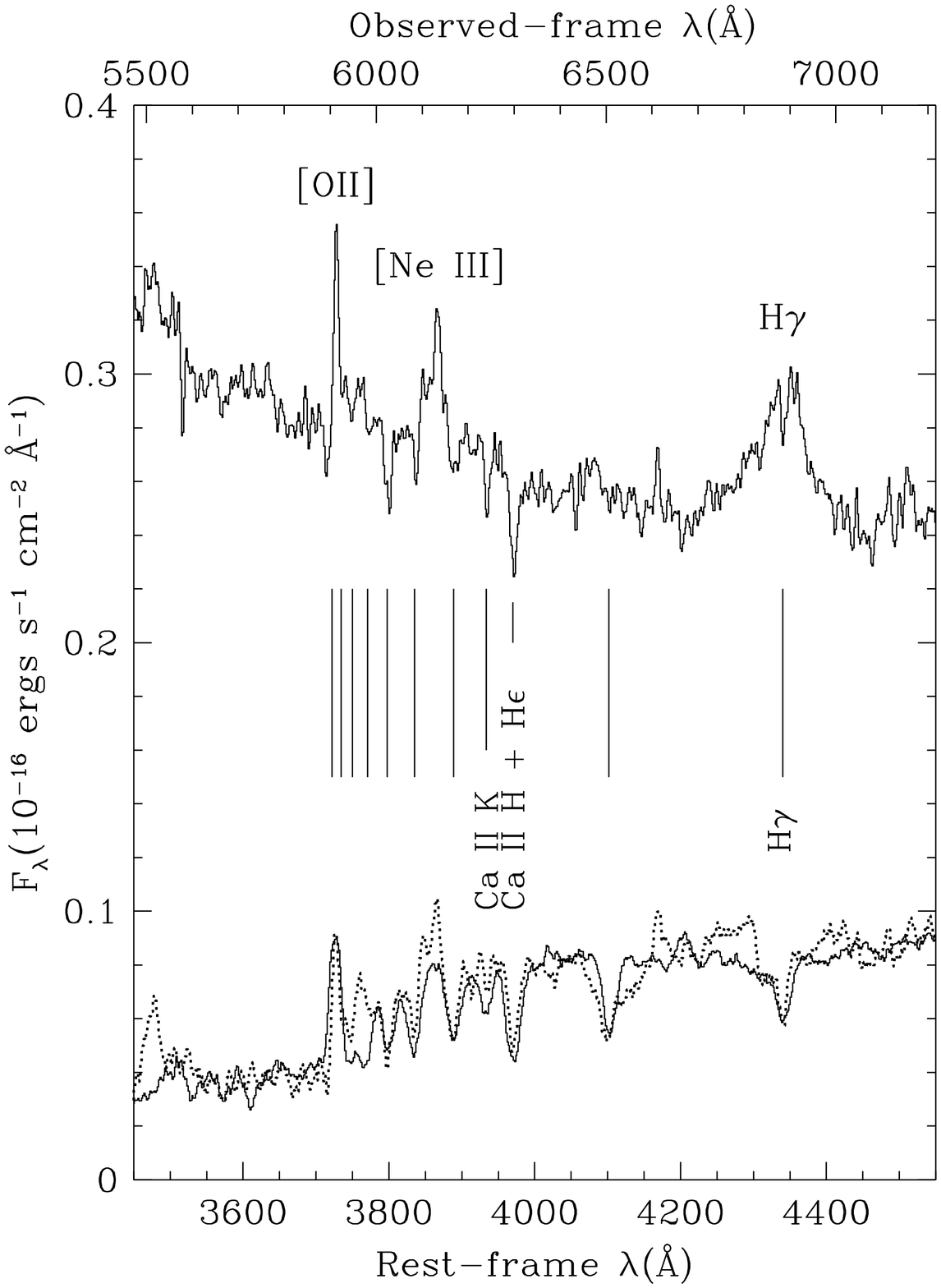,height=18cm}}
\figcaption{Top: The spectrum of quasar B, detail. 
Rest-frame wavelengths assume $z=0.5862$. 
Stellar absorption features are marked where expected for the
redshift, and include Ca II K lines and Balmer lines through H12.
At the bottom is shown the difference spectrum representative of the 
B host galaxy, as described in the text, and for comparison is a 
starburst galaxy FIRST J103540.0+355124 (White et al. 1999) (dotted line).} 

\end{document}